\documentclass{article}
\usepackage[utf8]{inputenc}
\usepackage{import}
\usepackage{amsmath}
\usepackage{slashed}
\usepackage{mathtools}
\usepackage{caption}
\usepackage{authblk}
\usepackage{blindtext}
\usepackage{subcaption}
\usepackage{color}
\usepackage{setspace}

\DeclarePairedDelimiter\ket{\lvert}{\rangle}
\DeclarePairedDelimiterX\braket[2]{\langle}{\rangle}{#1 \delimsize\vert #2}
\usepackage{abstract}
\usepackage{cite}
\usepackage{graphicx}
\title{\textbf{Phenomenology of $\tau^- \rightarrow \pi^- \nu_{\tau} \gamma$ using LCSR}}
\author[1,2]{Anshika Bansal\thanks{ anshika@prl.res.in}}
\author[1]{Namit Mahajan\thanks{ nmahajan@prl.res.in}}
\affil[1]{Physical Research Laboratory, Ahmedabad, 380009, India.}
\affil[2]{Indian Institute of Technology, Gandhinagar, 382424, India.}
\date{}
\begin{document}
\maketitle
\doublespacing
\begin{abstract}
    We present the study of radiative tau decay ($\tau^- \rightarrow \pi^- \nu_\tau \gamma$), computing the structure dependent contribution using Light Cone Sum Rules. This decay includes the same form factors as the radiative pion decay with the crucial difference that the momentum transfer squared, $t$, between the pion-photon system is positive, which makes these form factors timelike and also as $t$ can now take values upto $m_\tau^2$ , it can produce real hadronic resonances. 
 The analytical form for these form factors has been calculated using light cone sum rules method and the invariant mass-spectrum in the $\pi-\gamma$ system and the decay width are presented. The structure dependent parameter, $\gamma$, the ratio of the axial-vector to vector form factor is found to be in good agreement with the experimental determination. 
\end{abstract}
\section{Introduction}
$\tau$ is the heaviest lepton with $m_\tau = 1776.86 \pm 0.12 MeV$ \cite{pdg} and has numerous decay channels because of its heavy mass (see for example \cite{Barish:1987nj,Gentile:1995ue,Stahl:2000aq,Davier:2005xq,Pich:2013lsa} for different aspects of $\tau$ lepton physics.). It is the only lepton which can decay into hadrons. Theoretically, electroweak part is reasonably well established while one is still lacking in developing a proper methodology to understand the strong interactions. The study of hadronic $\tau$-decays helps us to understand the dynamics of strong interaction involved in the hadronisation of QCD currents in a cleaner environment. \\
In particular, we are interested in the study of radiative tau decay in the present work i.e. $\tau^- \rightarrow \pi^- \nu_\tau \gamma$. The branching ratio of 
$\tau^- \rightarrow \pi^- \nu_\tau$ is $(10.82\pm 0.05)\%$ \cite{pdg}. Hence, one expects the branching ratio for radiative tau decay to be $\mathcal{O}(10^{-3})$. To get a sense for this expectation, one can write the branching ratio as a product of branching ratios of $\tau\to\rho\nu_{\tau}$ and $\rho\to\pi\gamma$, and using the values from \cite{pdg}, one gets $\sim 10^{-3}$, which is about $10^{-2}$ of the non-radiative branching ratio. Even though the branching ratio is not very small, these decays are not observed experimentally yet which makes the study of this mode important. \\
 The decay amplitude of this process includes two contributions \cite{kim,volkov,Decker:1994dd,Geng:2003th,roig}:
\begin{itemize}
 \item \textbf{Internal Bremsstrahlung (IB):} The contribution coming from the emission from either the incoming or the outgoing particles. This contribution can be calculated trivially with the use of scalar QED for the point-like charged pion while the emission from the $\tau$ leg is calculated straightforwardly using QED. Diagramatically this is shown in (a) and (b) of Fig.(\ref{feyn}). 
 \item \textbf{Structure Dependent (SD):} This contribution is governed by the strong interactions and contains non-trivial parts. Pion can no longer be taken as a point-like particle. The partonic structure will play a role. This contribution appears because of the hadronisation of $J^P= 1^-$ ($\gamma^\mu$) and $1^+$($\gamma^\mu \gamma_5$) intermediate quark-antiquark currents of the matrix element ((c) of Fig.(\ref{feyn})) and hence depends on the long distance dynamics. Using the Lorentz symmetry, it can be parametrised by vector and axial-vector form factors $F_V^{(\pi)}$ and $F_A^{(\pi)}$, respectively. These form factors encode the information of strong dynamics involved in the hadronisation of these currents and their evaluation requires a non-perturbative treatment such as, Light Cone Sum Rules (LCSR), Chiral Perturbation Theory $\chi$PT or Lattice QCD. 
 SD contribution also includes a \textit{Contact Term (CT)}, which emerges as a consequence of gauge invariance and graphically represented in (d) of Fig.(\ref{feyn}).\\
 \begin{figure}
     \centering
     \includegraphics[width=9.5cm, height=7cm]{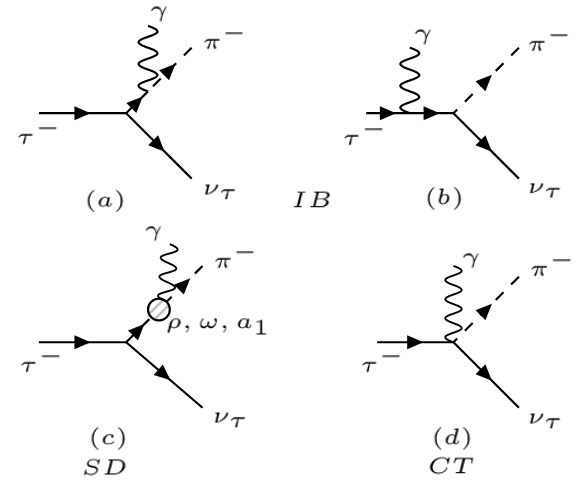}
     \caption{Feynman diagrams showing different contributions to the radiative tau decay. (a) and (b) represents the IB contribution, (c) represents the SD contribution and (d) represents the CT contribution. }
     \label{feyn}
 \end{figure}
\end{itemize}
The explicit form of these contributions will be calculated in Section-\ref{Amplitude} where we will see that the IB part consists of two contributions: one independent of $m_\tau$ and another proportional to $m_\tau$. The $m_\tau$ independent contribution turns out to be equal and opposite to the CT contribution and hence gets cancelled in the total amplitude. \\
The amplitude for the process of interest is related to that of the radiative pion decay by crossing symmetry with a major difference that comes at the level of kinematics as the square of the momentum transferred between the pion-photon and leptonic system can now take values up to $m_\tau^2$. While in the radiative  pion decay, it can take values only upto $m_\pi^2$ which is almost negligible. Also, as both pion and photon are in the final state, the form factors involved in this process are time-like, and hence complicated, unlike the form factors involved in the radiative pion decay which are space-like. As a consequence, the light flavoured mesons will be created on-shell and give resonant structures in the pion-photon invariant mass spectrum.  \\
Hence to understand this process, the main task is to calculate the time-like form factors involved in the process. These form factors probe the structure of the pion. The information about the pion structure can be obtained by determining the ratio of $F_A^{(\pi)}(0)$ to $F_V^{(\pi)}(0)$ which is defined as the structure dependent parameter, $\gamma$  i.e. $\gamma = \frac{F_A^{(\pi)}(0)}{F_V^{(\pi)}(0)}$. We know the values of $F_A^{(\pi)}(0)$ and $F_V^{(\pi)}(0)$ from the experimental determination of radiative pion decay to be equal to ($0.0119\pm0.0001$) and ($0.0254\pm 0.0017$), respectively\cite{pdg}, which results into the value of $\gamma$ equal to ($0.4685\pm0.0353 $). The value of $\gamma$, which is the ratio of form factors evaluated at zero momentum transfer, will be same for radiative tau and pion decays. The calculation of radiative tau decay helps in determining this structure dependent parameter theoretically in a consistent way. This decay is also useful to understand the light-by-light hadronic contribution to the muon anomalous magnetic moment, $(g-2)_\mu$ \cite{Cirigliano:2002pv}. In \cite{sree}, the authors have discussed how this decay can provide the means for the tau neutrino mass determination. These gauge invariant form factors for the radiative tau decay have been parametrised using Breit-Wigner type resonances \cite{decker}, light front quark model \cite{Geng:2003th} and resonance $\chi$PT \cite{roig} in the past. \\
The differences in the literature stem from the vastly different approaches adopted to determine or estimate the form factors, which affect the predictions for the rate and spectrum, as well as extraction of $\gamma$, including the sign. As an example, whenever the resonances are included via Breit-Wigner method, a suspecting issue always is the relative phase between the different contributions. The main aim of this paper is to calculate these form factors using the method of LCSR in a consistent way.

The rest of the paper is organised as follows; in Section-\ref{Amplitude}, we present the amplitude calculation for the process and explicitly write the forms of different contributions mentioned above.
In Section-\ref{FF}, we present the calculation of the form factors using the method of LCSR and in Section-\ref{results} we report our results. Finally, in Section-\ref{discussion} we conclude our results with some remarks. Various definitions and conventions used are reported in Appendix-\ref{appendixA}. The values of various parameters used for numerical calculation are collected in Appendix-\ref{appendixB} and the kinematical details are provided in Appendix-\ref{kinematics}.

\section{Amplitude Computation}
\label{Amplitude}
Photon can be emitted by any charged particle. Hence in the present case, the photon can be emitted from either the pion or tau-lepton, 
as shown in Fig.(\ref{feyn}). The pion is a composite object with quark-antiquark pair. 
Therefore, the internal structure of the pion will also contribute to the process. This gives rise to two non-perturbative Form Factors.
As mentioned above, the amplitude of radiative tau decay includes various contributions: 
Internal Bremsstrahlung (IB), Structure Dependent (SD) and Contact term (CT). IB contribution comes from the emission of 
the photon from tau and pion (considering pion to be the point object). SD contribution comes from the emission of photon from
the internal structure of the pion. The contact term in an interesting effective contribution and has its origin in the gauge 
invariance of a QED amplitude \cite{countingCT}. We follow this approach here.\\
The amplitude of the process $\tau^-(p_1) \rightarrow \pi^-(p_2) \nu_\tau(p_3) \gamma(k)$ can be written as 
(employing the low energy four-Fermi effective Hamiltonian obtained by integrating out the heavy W-boson),
\begin{equation}
 \mathcal{A}(\tau^- \rightarrow \pi^- \nu_\tau \gamma) = \frac{G_F}{\sqrt{2}} V_{ud} \left<\pi^- \nu_\tau \gamma| (\bar \nu_\tau \Gamma^\mu \tau )(\bar d \Gamma_\mu u )|\tau^- \right>
\end{equation}
where, $\Gamma^\mu = \gamma^\mu (1-\gamma_5)$.\\
This amplitude can be factorised in two parts; one where the photon is emitted from the final state pion and another where the photon 
gets emitted from the initial state tau lepton. 
\begin{align}
    \mathcal{A}(\tau^- \rightarrow \pi^- \nu_\tau \gamma) \nonumber &= \frac{G_F}{\sqrt{2}} V_{ud} \left[\left<\pi^- \gamma|(\bar d \Gamma_\mu u )|0\right>\left<\nu_\tau|(\bar \nu_\tau \Gamma^\mu \tau)|\tau^-\right>  +   \left<\nu_\tau \gamma|(\bar \nu_\tau \Gamma^\mu \tau)|\tau^-\right>\left<\pi^-|(\bar d \Gamma_\mu u)|0\right>\right] \\ 
    &= \frac{G_F}{\sqrt{2}} V_{ud} \left[- i e \epsilon^*_\alpha(\bar u_\nu \Gamma_\mu u_\tau)\int d^4x e^{ikx} \left<\pi^-|T\{j_{em}^\alpha (x)\bar d \Gamma^\mu u(0)\}|0\right> \right. \nonumber \\ & \hspace{2cm}- \left. ef_\pi p_{2\mu} \epsilon^*_\alpha \int d^4 x e^{ikx} \left<\nu_\tau|T\{j_{em}^\alpha (x) \bar \nu_\tau \Gamma^\mu \tau(0) \}|\tau^-\right>\right]
 \label{amp}
\end{align}
where,  $j_{em}^\alpha(x) = Q_\psi \bar\psi(x)\gamma^\alpha \psi(x) = - \bar{\tau}\gamma^\alpha \tau + Q_u \bar{u}\gamma^\alpha u + Q_d \bar{d}\gamma^\alpha d$ and $f_\pi$ is the pion decay constant. 
The conventions and definitions are given in Appendix-\ref{appendixA}.  
This factorization of the amplitude holds for energetic photons and at the leading order in $\frac{1}{m_\tau}$ and $\alpha_{em}$.\\
For the computation of the first term of Eq.(\ref{amp}), define the hadronic matrix element as,
\begin{equation}
 T^{\alpha \mu}(p_2,k) = i\int d^4 x e^{ikx} \left<\pi^-|T\{j_{em}^\alpha(x) \bar d \Gamma^\mu u(0)\}|0\right>.
 \label{talpha}
\end{equation}
Using the conservation of electromagnetic current, one can apply the Ward identity which results into 
\begin{align}
  k_\alpha T^{\alpha \mu}(p_2,k) &= \left<\pi^-|\bar d(0) \Gamma^\mu u(0)|0\right>  = if_\pi p_2^\mu
 \label{ward}
\end{align}
in the momentum space.\\
Also, one can write the hadronic matrix element (defined in Eq.(\ref{talpha})) using the general covariant decomposition in terms of the pion and photon momentum i.e. $p_2$ and $k$ respectively, as
\begin{align}
 T^{\alpha \mu}(p_2,k) = A g^{\alpha \mu} + B p^{2\alpha}p^{2\mu} + C p^{2\alpha}k^\mu +D k^\alpha p^{2\mu} + E k^\alpha k^\mu + iF_V^{(\pi)} \epsilon^{\alpha \mu\beta\nu} p_{2\beta} k_\nu
 \label{genpar}
\end{align}
where, $A,B,C,D,E,F_V^{(\pi)}$ are gauge invariant scalar functions of $(p_2+k)^2$. 
Contraction of Eq.(\ref{genpar}) with $k_\alpha$, results in (for on-shell photon $k^2 = 0$ and Levi-civita tensor is anti-symmetric in $\alpha$ and $\nu$): 
\begin{equation}
 k_\alpha T^{\alpha\mu}(p_2,k)= A k^\mu +B(p_2.k)p_2^{\mu}+C(p_2. k)k^\mu.
 \label{Talphamu}
\end{equation}
On equating Eq.(\ref{ward}) and Eq.(\ref{Talphamu}), we get
\begin{equation}
     C = \frac{-A}{(p_2.k)}, \hspace{0.5cm}\text{    and}\hspace{2cm} B= \frac{if_\pi}{(p_2.k)}
\end{equation}
which results in the final form of hadronic matrix element to be,
\begin{equation}
  T^{\alpha\mu}(p_2,k) = F_A^{(\pi)} \left[g^{\alpha\mu}(P.k)-P^\alpha k^\mu\right] + iF_V^{(\pi)} \epsilon^{\alpha\mu\beta\nu} P_\beta k_\nu - if_\pi g^{\alpha\mu} + if_\pi \frac{P^\alpha P^\mu}{P.k}.
  \label{ff}
\end{equation}
Here, $F_A^{(\pi)} =\frac{A+if_\pi}{P.k} $  and  $P= p_1-p_3 = p_2+k$ and $p_2.k = P.k$. Hence, the first term in Eq.(\ref{amp}) reads,
\begin{align}
 \left<\pi^- \gamma|\bar d \Gamma_\mu u|0\right>\left<\nu_\tau|\bar \nu_\tau \Gamma^\mu \tau|\tau^-\right> \nonumber &= ie \epsilon^{*\alpha} \left[\bar u_\nu \Gamma^\mu u_\tau \right]\left[iF_A^{(\pi)}\left\{g_{\alpha\mu}(P.k)-P_\alpha
 k_\mu\right\}-F_V^{(\pi)}\epsilon_{\alpha\mu\beta\nu}P^\beta k^\nu\right] \\ &+ ie \epsilon^{*\mu} f_\pi \bar u_\nu \Gamma_\mu u_\tau - ie f_\pi \frac{\epsilon^*.P}{P.k} \bar u_\nu \slashed{P} (1-\gamma_5) u_\tau .
 \label{1stf}
\end{align}
The second term in Eq.(\ref{amp}), using QED Feynman rules, takes the form
\begin{align}
 \left<\nu_\tau \gamma|\bar \nu_\tau \Gamma^\mu \tau|\tau^-\right>\left<\pi^-|\bar d \Gamma_\mu u|0\right>   &= -ief_\pi \bar u_\nu(p_3) \slashed{\epsilon}^*(1-\gamma_5)u_\tau(p_1) \nonumber \\  &+\frac{ief_\pi m_\tau}{2p_1.k} \left\{ \bar u_\nu (p_3) \left[(2\epsilon^* .p_1)- \slashed{k}\slashed{\epsilon}^*\right] (1+\gamma_5)u_\tau(p_1)\right\}.
 \label{2nd}
\end{align}
Adding the two, the final form of the amplitude is:
\begin{align}
 \mathcal{A}(\tau^- \rightarrow \pi^- \nu_\tau \gamma)&= \frac{G_F}{\sqrt{2}} V_{ud} \left[ie\epsilon^{*\alpha}(\bar u_\nu \Gamma^\mu u_\tau)\left\{iF_A^{(\pi)}\left[g_{\alpha\mu}(P.k)-P_\mu k_\alpha\right]-F_V^{(\pi)}\epsilon_{\alpha\mu\beta\nu} P^\beta k^\nu\right\} \right. \\ \nonumber &+ \left. ief_\pi m_\tau \bar u_\nu \left\{\frac{\epsilon^*.p_1}{p_1.k}- \frac{ \slashed{k}\slashed{\epsilon
}^*}{2 p_1.k} - \frac{\epsilon^*.p_2}{p_2.k}\right\}(1+\gamma_5) u_\tau \right].
\label{amplitude}
\end{align}
Here, $F_A^{(\pi)}$ and $F_V^{(\pi)}$ are the gauge invariant axial-vector and vector form factors, respectively. The contact term appears explicitly by the use of Ward identity and cancels against the $m_\tau$ independent contribution of photon emission from $\tau$. \\
For further simplification, we have divided the full amplitude as,
\begin{equation}
     \mathcal{A}(\tau^- \rightarrow \pi^- \nu_\tau \gamma)  = \mathcal{A}_{IB} + \mathcal{A}_{V}+\mathcal{A}_A  \hspace{0.2cm}=  \mathcal{A}_{IB} + \mathcal{A}_{SD}.
 \label{final}
\end{equation}
Here,
\begin{equation}
 \mathcal{A}_{IB} = \frac{G_F}{\sqrt{2}} V_{ud}\left[ief_\pi m_\tau \bar u_\nu \left\{\frac{\epsilon^*.p_1}{p_1.k}- \frac{ \slashed{k}\slashed{\epsilon
}^*}{2 p_1.k} - \frac{\epsilon^*.p_2}{p_2.k}\right\}(1+\gamma_5) u_\tau \right],
\end{equation}
\begin{equation}
 \mathcal{A}_{V} = -\frac{G_F}{\sqrt{2}} V_{ud} \left[ie\epsilon^{*\alpha}(\bar u_\nu \Gamma^\mu u_\tau)\left(F_V^{(\pi)}\epsilon_{\alpha\mu\beta\nu} P^\beta k^\nu\right) \right], \text{ and}
\end{equation}
\begin{equation}
\mathcal{A}_{A} = \frac{G_F}{\sqrt{2}} V_{ud} \left[ie\epsilon^{*\alpha}(\bar u_\nu \Gamma^\mu u_\tau)\left(iF_A^{(\pi)}\left[g_{\alpha\mu}(P.k)-P_\mu k_\alpha\right]\right)\right].
\end{equation}
$\mathcal{A}_{V}$ and $\mathcal{A}_{A} $ combined gives the structure dependent contribution, while $\mathcal{A}_{IB}$ is the  
internal bresstrahlung contribution. \\

\section{Form Factors in LCSR framework}
\label{FF}
In the previous section, we saw that the amplitude of the radiative tau decay depends on two gauge invariant form factors; $F_A^{(\pi)}$ and $F_V^{(\pi)}$. These form factors are the non-perturbative objects and need a non-perturbative treatment. In this section, we will calculate these form factors using the method of LCSR.\\
The method of sum rules was developed in 1979 by Shifman, Vainshtein and Zakharov (SVZ) \cite{Shifman:1978bx,Shifman:1978by}. Their basic idea was to use the analytic properties of a correlation function (treated in the framework of operator product expansion (OPE)) to derive the hadronic parameter involved in a process. Below we  briefly outline the method (for details, see \cite{khodjamirian,braun,Shifman:1998rb}).\\
The important tools for deriving the sum rules are: dispersion relation, operator product expansion(OPE), quark-hadron duality and the Borel transformation. The dispersion relation relates the real part of the correlation function to its imaginary part using the Cauchy's integral formula. According to OPE, the correlation function can be written as a sum of products of long distance matrix elements of operators of increasing dimension and short distance Wilson coefficients which can be calculated using perturbation theory. The higher dimension operators capture the information of QCD vacuum fields in the form of vacuum condensates. Both dispersion relation and OPE gives the same physics and hence can be equated.\\
Operationally, quark hadron duality means,
\begin{equation}
    q^2 \int_{s_0^h}^\infty ds \frac{\rho^h(s)}{s(s-q^2)} \simeq \frac{q^2}{\pi} \int_{4m^2}^\infty ds \frac{Im \Pi^{(pert)}(s)}{s(s-q^2)}.
\end{equation}
%if $Im \Pi(s) \rightarrow Im \Pi^{pert}(s)$ at $s\rightarrow+\infty$. 
Here, $\rho^h$ is the hadronic spectral density function, while  $\Pi^{pert}(s)$ (or $\Pi^{QCD}(s)$) is the perturbatively calculated correlation function. We will use this duality approximation below.\\
As the correlation function has contribution from all the resonance states as well as the continuum, one performs Borel transformation to suppress the effect of higher resonances and continuum. Mathematically, the Borel transform is given by,
\begin{equation}
    \Pi(M^2) \equiv \mathcal{B}_{M^2} \Pi(k^2) = \lim_{-k^2, n \rightarrow \infty , -k^2/n = M^2} \frac{(-k^2)^{(n+1)}}{n!} \left(\frac{d}{dk^2}\right)^n \Pi(k^2),
\end{equation}

where M is known as the Borel parameter.\\
It was noticed that these SVZ sum rules have some limitations such as: the OPE upsets the power counting in large $Q^2$ and that even after performing the Borel transformation, practical calculations suffer from unsuppressed contributions. 
  These limitations can be overcome by using light cone sum rules (LCSR).
In LCSR, one expands the products of the currents near the light cone. LCSR give vacuum-to-hadron correlation function while by SVZ sum rules one get vacuum-to-vacuum correlation functions. In LCSR, OPE at short distances is replaced by systematic expansion in the transverse direction in infinite momentum frame.\\
In the lightcone limit, the bi-local operator sandwiched between the pion state and vacuum is expressed as,
\begin{equation}
    \left<\pi^0(p)|\bar u(y) \gamma_\mu \gamma_5 u(x)|0\right>_{x^2=0} = -if_\pi p_{\mu }\int_0^1 du \hspace{0.1 cm} e^{i(up_2.y+\bar u p.x)}\phi(u,\mu)
\end{equation}
   where, $\bar u = 1-u$ and $\phi(u,\mu)$ is leading twist-2 distribution amplitude given by 
   \begin{equation}
       \phi_\pi(u,\mu) = 6 u \bar u \left[1+\sum_{n=2,4\ldots} a_n(\mu)C_n^{3/2}(u-\bar u)\right].
       \label{phi}
   \end{equation}
   Here, $C_n^{3/2}$ are the Gegenbauer polynomials and $a_n$ are the multiplicatively renormalisable coefficient defined as,
   \begin{equation}
       a_n(\mu) = a_n(\mu_0) \left( \frac{\alpha_s(\mu)}{\alpha_s(\mu_0)} \right)^{\gamma_n/\beta_0}
       \label{a2}
   \end{equation}
 with $ \alpha_s = \frac{g_s^2}{4\pi} $ ($g_s$ is the strong coupling constant), $\beta_0$ is the leading QCD $\beta$-function and 
 \begin{equation}
     \gamma_n = \frac{4}{3}\left[-3-\frac{2}{(n+1)(n+2)}+4\left(\sum_{k=1}^{(n+1)}\frac{1}{k}\right)\right].
 \end{equation}

The remaining process for computation is same as for SVZ sum rules. We are now ready to derive the form factors, $F_V^{(\pi)}$ and $F_A^{(\pi)}$ using this technique.\\
As we know, these form factors arises from the computation of the hadronic matrix element defined in Eq.(\ref{talpha}), i.e.,
\begin{equation}
 T^{\alpha\mu}(p_2,k) = i\int d^4 x e^{ikx} \left<\pi^-|T\left\{Q_u \bar u \gamma^\alpha u(x)\bar d \Gamma^\mu u(0) + Q_d \bar d \gamma^\alpha d(x) \bar d \Gamma^\mu u(0)\right\}|0\right>
 \label{ta}
\end{equation}
where, $Q_u$ and $Q_d$ are the charges of up and down quark respectively in units of $e$.
Using the definitions and identities given in Appendix-\ref{appendixA}, we get
\begin{align}
 T^{\alpha\mu}(p_2,k)  \nonumber &= i f_\pi \int d^4 x \frac{e^{ikx}}{2 \pi^2 x^4}\int_0^1 d u  \phi (u,\mu) \left[i\epsilon^{\mu\beta\alpha\rho}x_\beta p_{2\rho}\left(Q_u e^{i\bar u p_2 x}+Q_d e^{iup_2x}\right)\right. \\ &+ \left. \left(x^\mu p_{2}^\alpha-g^{\mu\alpha}(x.p_2)+x^\alpha p_{2}^\mu\right) \left(Q_u e^{i\bar u p_2 x}-Q_d e^{iup_2x}\right) \right],
\end{align}
where, as mentioned above, $\phi(u,\mu)$ is the pion distribution amplitude and $\bar u = 1-u$. The integration over $x$ results into,
\begin{align}
 T^{\alpha\mu}(P,k) \nonumber&= i f_\pi  \left[\frac{i\epsilon^{\mu\beta\alpha\rho} }{3}P_{\rho} k_\beta \int_0^1 du \frac{\phi(u,\mu)}{P^2 \bar u + k^2 u} + 2 \left\{P^\alpha P^\mu- (P.k) g^{\mu \alpha}\right\} \int_0^1 du \frac{\phi(u,\mu) \bar u}{P^2 \bar u + k^2 u}\right. \\  & -\left.  \left\{g^{\mu\alpha}(P.k) - P^\alpha k^\mu\right\}\left\{ \int_0^1 du \phi(u,\mu)\left(\frac{1-2 \bar u}{P^2 \bar u + k^2 u}\right)\right\}  \right]   .
 \label{lightcone}
\end{align}
Here, $p_2+k = P $ and we have used the fact that the distribution amplitude is a symmetric function of $u$ and $\bar u$.\\
A comparison with the general decomposition of the hadronic tensor given in Eq.(\ref{ff}) yields the following forms of vector and axial-vector form factors from QCD calculation.
\begin{equation}
 F_V^{QCD}(t) = \frac{i f_\pi}{3}\int_0^1
du \frac{\phi(u,\mu)}{t \bar u + k^2 u}  \hspace{1cm} 
\label{FV}
\end{equation}
$$\implies \frac{1}{\pi}Im\{F_V^{QCD}(t)\} = \frac{i f_\pi}{3}\int_0^1
du \phi(u,\mu) \delta(t \bar u + k^2 u), \text{ and}$$ 
\begin{equation}
 F_A^{QCD}(t) = -i f_\pi \int_0^1
du \phi(u,\mu)\left(\frac{1-2 \bar u}{t \bar u + k^2 u}\right)
\label{FA}
\end{equation}
$$\implies \frac{1}{\pi}Im\{F_A^{QCD}(t)\} = -i f_\pi\int_0^1
du \phi(u,\mu) (1-2 \bar u)\delta(t \bar u + k^2 u).$$
Here, $t\equiv P^2=(p_2+k)^2 = (p_1-p_3)^2$ is the invariant mass square of the photon-pion system.\\
%\subsection{Dispersion relation and Sum rule}
Now, after computing the perturbative QCD contribution, the analytic properties of this hadronic matrix element are used to 
derive the contribution of various hadronic states. It will get contribution from ($\rho,\omega,a_1$-mesons)$+$ higher resonances 
and the continuum. In the present case, contributions coming from $\rho,\omega,a_1$-mesons will saturate the sum rules 
and thus will be the focus here.\footnote{ The contribution of the higher resonances, at the present level of accuracy, is roughly 20\% of these 
resonances because of the Borel suppression.}\\
 
Considering the matrix element $\left<\pi^-|T\{j_{em}^\alpha(x)j_{ew}^\mu(0)\}|0\right>$ and inserting a complete set of states, we get,
\begin{equation}
 \left<\pi^-|T\{j_{em}^\alpha(x)j_{ew}^\mu(0)\}|0\right> = \left<\pi^-|j_{em}^\alpha(x)|n\right>\left<n|j_{ew}^\mu(0)|0\right>
\end{equation}
where, $\ket n = \ket \rho + \ket \omega +  \ket {a_1} + $ higher resonances + continuum.\\
\begin{itemize}
 \item \textbf{$\rho$ and $\omega$-meson} contribution:
 The $\rho$-meson contribution will come from,
\begin{equation}
 \left<\pi^-(p_2)|j_{em}^\alpha(x)|\rho(p_2+k)\right>\left<\rho(p_2+k)|j_{ew}^\mu(0)|0\right>.
\end{equation}
Using the definitions given in Appendix-\ref{appendixA}, 
\begin{equation}
  \left<\pi^-(p_2)|j_{em}^\alpha(x)|\rho(p_2+k)\right>\left<\rho(p_2+k)|j_{ew}^\mu(0)|0\right> = i m_\rho f_\rho \epsilon^{\alpha \lambda\beta\nu} g^\mu_\lambda p_{2\beta}k_\nu F_{\rho\pi}(k^2)
\end{equation}
where, $m_\rho$ and $f_\rho$ are the mass and decay constant of $\rho$-meson respectively. Neglecting the very small difference between the masses of $\rho$ and $\omega$, the contribution of $\omega$ will be equal to the contribution of $\rho$ and hence multiplying $\rho$ contribution by a factor of $2$ will incorporate the contribution of $\omega$-meson.
\item \textbf{$a_1$-meson contribution:}
The $a_1$-meson contribution will come from,
\begin{equation}
 \left<\pi^-(p_2)|j_{em}^\alpha(x)|a_1(p_2+k)\right>\left<a_1(p_2+k)|j_{ew}^\mu(0)|0\right>,
\end{equation}
which results into
\begin{equation}
  \left<\pi^-(p_2)|j_{em}^\alpha(x)|a_1(p_2+k)\right>\left<a_1(p_2+k)|j_{ew}^\mu(0)|0\right> = i m_{a_1} f_{a_1} \left[2p_2.k g^{\alpha\mu}-2p_2^\alpha k^\mu\right] G_{a_1\pi}(k^2)
\end{equation}
using the definitions given in Appendix-\ref{appendixA}.
Here, $m_{a_1}$ and $f_{a_1}$ are the mass and decay constant of $a_1$-meson respectively.
\end{itemize}
Here, $F_{\rho\pi}$($G_{a_1\pi}$) captures the physics of transition of $\rho$($a_1$)-meson to the $\pi$-meson.
Using the optical theorem in Eq.(\ref{talpha}), we get,
\begin{equation}
    2 Im\{T^{\alpha\mu}(p_2,k)\} = \sum_n \left<\pi^-|j_{em}^\alpha(x)|n\right> \left<n|j_{ew}^\mu|0\right>d\tau_n (2\pi)^4 \delta^4(k-p_n),
\end{equation}
and from Cauchy's theorem,
\begin{equation}
    T(k^2)= \frac{1}{\pi}\int_{t_{min}}^\infty ds \frac{Im\{T(s)\}}{s-k^2-i\epsilon}.
\end{equation}
Substituting the contributions of $\rho$ and $a_1$, we get,
\begin{align}
 T^{\alpha\mu}(p_2,k) \nonumber & = \frac{2i m_\rho f_\rho \epsilon^{\alpha \lambda\beta\nu} g^\mu_\lambda p_{2\beta}k_\nu F_{\rho\pi}(k^2)}{m_\rho^2-(p_2+k)^2-im_\rho \Gamma_\rho} + \frac{ i m_{a_1} f_{a_1} \left[2p_2.k g^{\alpha\mu}-2p_2^\alpha k^\mu\right] G_{a_1\pi}(k^2)}{m_{a_1}^2-(p_2+k)^2-im_{a_1} \Gamma_{a_1}} \\   &+ \frac{1}{\pi}\int_{s_0^h}^\infty ds \frac{Im\{T^{\alpha\mu}(s,k)\}}{s-k^2-i\epsilon}.
 \label{dispersion}
\end{align}
Here, $s_0^h$ is the threshold of the lowest continuum state and $\Gamma_\rho$ and $\Gamma_{a_1}$ are the decay widths of $\rho$ and $a_1$ mesons, respectively. 
This is the dispersion relation which relates the imaginary part to the real part. Now, the light cone sum rules can be derived by taking the form of $F_V^{(\pi)}(t)$ from this dispersion relation and equating it with the form obtained in Eq.(\ref{FV}), i.e.  
\begin{equation}
 \frac{2m_\rho f_\rho F_{\rho\pi}(k^2)}{m_\rho^2-t-im_\rho \Gamma_\rho} +  \frac{1}{\pi}\int_{s_0^h}^\infty ds \frac{Im\{F_V(s)\}}{s-t-i\epsilon} = \frac{if_\pi}{3}\int_0^1 du \frac{\phi(u,\mu)}{t \bar u + k^2 u}.
\end{equation}
Using the duality approximation and the Chauchy's integral, 
\begin{equation}
 \frac{1}{\pi}\int_{s_0}^\infty ds \frac{Im\{F_V(s,k)\}}{s-t-i\epsilon} =\frac{1}{\pi}\int_{s_0^\rho}^\infty ds \frac{Im\{F_V^{QCD}(s,k)\}}{s-t-i\epsilon} = \frac{if_\pi}{3}\int_{u_0}^1 du \frac{\phi(u)}{t \bar u + k^2 u},
\end{equation}
with $u_0 = \frac{s_0}{k^2+s_0} = 1$ (as $k^2=0$).
As a result, the sum rule for $F_V^{(\pi)}(t)$ turns out to be,
\begin{equation}
 \frac{2 m_\rho f_\rho F_{\rho\pi}(k^2)}{m_\rho^2-t} = \frac{if_\pi}{3}\int_0^{u_0} du \frac{\phi(u)}{t \bar u + k^2 u}.
\end{equation}
Similarly, by equating the form of $F_A^{(\pi)}(t)$ obtained from the dispersion relation with the form given in Eq.(\ref{FA}) and using the duality approximation, the sum rule for $F_A^{(\pi)}(t)$ reads,
\begin{equation}
 \frac{2 i m_{a_1}f_{a_1}G_{a_1\pi}(k^2)}{m_{a_1}^2-t} = -i f_\pi \int_0^{u_0} \phi(u) \left(\frac{1-2 \bar u}{t \bar u + k^2 u}\right).
\end{equation}
After Borelisation and substituting these sum rules back in Eq.(\ref{dispersion}), we get the following analytical forms for $F_V^{(\pi)}$ and $F_A^{(\pi)}$,
\begin{equation}
 F_V^{(\pi)}(t) = -i\frac{f_\pi}{3 (m_\rho^2-t-i m_\rho \Gamma_\rho)}\int_0^1 du \frac{\phi(u)}{ \bar u }e^{\frac{m_\rho^2}{M^2}},
 \label{FVF}
\end{equation}
\begin{equation}
 F_A^{(\pi)}(t) = -i\frac{f_\pi }{m_{a_1}^2-t-i m_{a_1} \Gamma_{a_1}}\int_0^1  \frac{\phi(u)}{ \bar u }(1-2 \bar u)e^{ \frac{m_{a_1}^2}{M^2}}.
 \label{FAF}
\end{equation}
Here, $M$ is the Borel parameter and we have used the on-shell condition for photon (i.e. $k^2=0$).\footnote{ It is to be noted that these form factors have dimension of inverse mass and there is an extra factor of $-i$ due to the way initial amplitude is defined: $\mathcal{A}(\tau^- \rightarrow \pi^- \nu_\tau \gamma)$ instead of $i \mathcal{A}(\tau^- \rightarrow \pi^- \nu_\tau \gamma)$ as is often done.}

For the present calculation, we will use the asymptotic form (where $\mu\rightarrow \infty$) and Chernyak-Zhitnisky form (where $C_2$ term will be considered) of the pion distribution amplitude given in Eq.(\ref{phi}). Explicitly these forms are given by,
\begin{equation}
    \phi_\pi^{asym} (u,\mu) = 6 u \bar u, \text{ and}
    \label{phia}
\end{equation}
 \begin{equation}
     \phi_\pi^{CZ}(u,\mu) = 6 u \bar u\left[1+\frac{3a_2(\mu)}{2}\{5(u-\bar u)^2 -1\}\right]
     \label{phiz}
 \end{equation}
where, $a_2(\mu)$ is defined in Eq.(\ref{a2}) with $n=2$.
All the structure dependent information of the pion involved in the radiative tau decay is contained in the ratio of axial vector form factor and the vector form factor at zero invariant mass square of the photon-pion system, i.e.,
\begin{equation}
    \gamma = \frac{F_A^{(\pi)}(0)}{F_V^{(\pi)}(0)}
    \label{sdp}
\end{equation}
where, $\gamma$ is known as the structure dependent parameter (SDP). 
The vector form factor at $t=0$ can be related to the anomaly term (or Wess-Zumino-Witten term) in 
$\pi\gamma\gamma$ vertex ($1/(4\pi^2 f_{\pi})$). 
Using what is referred to as KSFR-II relation (\cite{Kawarabayashi:1966kd,Riazuddin:1966sw}), 
$m_{\rho}^2=2g_{\rho\pi\pi}^2f_{\pi}^2$, along with the assumptions of universality of $\rho$-coupling 
($g_{\rho\pi\pi}=g_{\rho NN}=g_{\rho\gamma}=g=2\pi\sqrt{3/N_c}$) and $\rho$ meson dominance of the pion electromagnetic form factor, 
one finds the right form emerging from $F_V^{(\pi)}(0)$, up to the overall factor $e^{\frac{m_\rho^2}{M^2}}$ which should tend to unity.
As we see later, the choice of the Borel parameter that provides a stable window, trivially yields unity for this factor within a few percent. \\
Before discussing the results, it may be worthwhile to ponder over possible duality violations. Such contributions arise from our use
of perturbatively evaluated spectral functions, imaginary parts of the form factors here, over the entire kinematical range. It is
notoriously difficult to exactly quantify the magnitude of such duality violating terms. However, it is rather important to have some
estimate or an educated guess since these would otherwise cause large uncertainties in the final results. For the case at hand,
the perturbative effects occus at $1/Q$, where hard scale $Q\sim m_{\tau}$ while the time scale over which the partons come together to form
final hadrons $\sim Q/\Lambda^2_{QCD}$. One possible way to evaluate the duality violations could be to use an instanton model, where the
light quark amplitudes will be suppressed. A rough calculation yields a quantity that in the Euclidean domain has the form
$Exp[-Q\rho]/Q^n$, where $\rho$ denotes the mean instanton size. Analytically continued to the Minkowski space, this would have an
oscillating factor multipled by negative powers of the energy released in the hard process $m_{\tau}$. Alternatively, one could assume
a comb of hadronic resonances that would contribute and carry out the algebra. Both lead to similar conclusions that the violations
are $\sim 10\%$ \cite{Shifman:2000jv} (also see \cite{GonzalezAlonso:2010rn,Dominguez:2016xhu} for detailed analyses for inclusive tau decays). This is the
typical duality violation contribution that we expect, though a more detailed calculation can reveal the actual amount of such violations.

\section{Results}
\label{results}
\begin{figure}
     \centering
     \includegraphics[width=9.5cm, height=6.5cm]{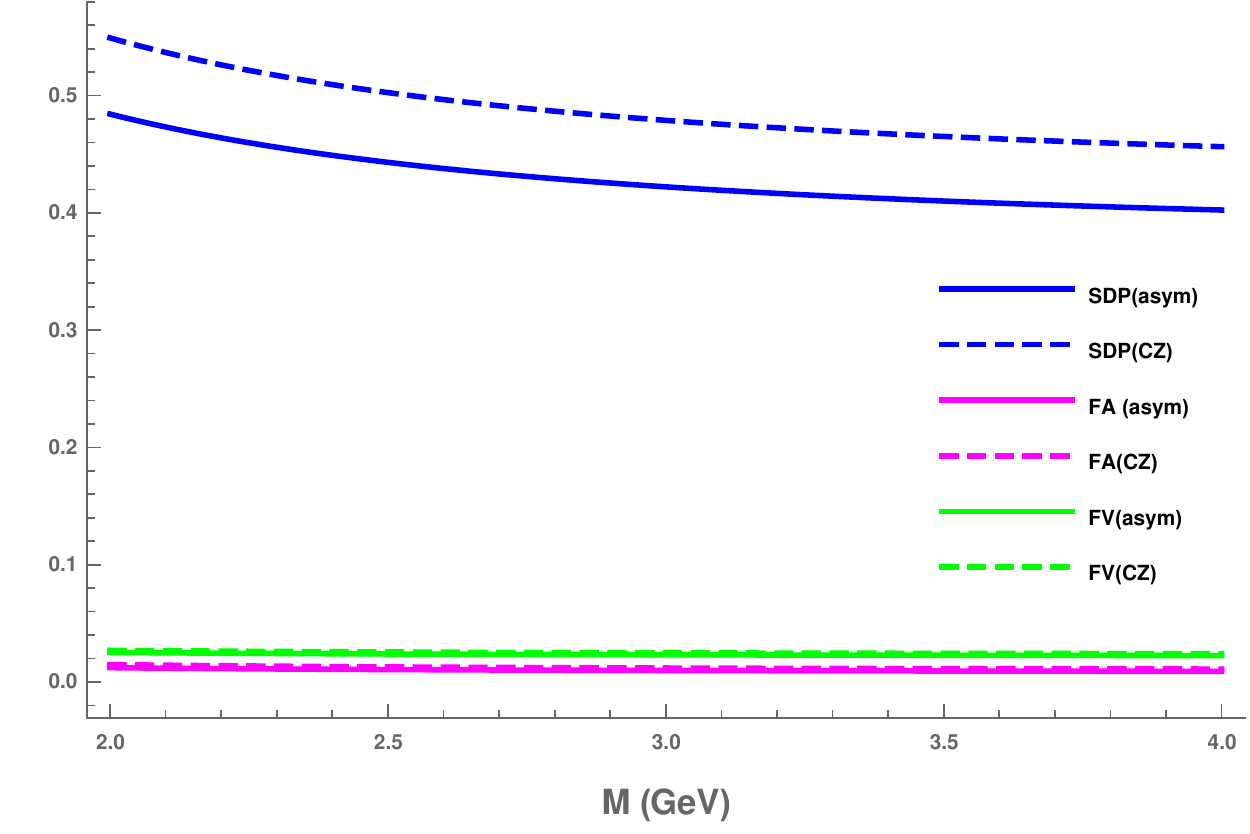}
     \caption{The dependence of structure dependent parameter (SDP), $F_A^{(\pi)}(0)$ and $F_V^{(\pi)}(0)$ on the Borel parameter $M$ (in GeV units) is shown in Blue, Magenta and Green, respectively. In this plot, form factors have been multiplied by $i m_{\pi}$ to make them dimensionless in and take care of the extra $-i$ in the form factors as noted in the Footnote1.}
     \label{sdpp}
 \end{figure}

\begin{figure}
\centering
\begin{subfigure}{.5\textwidth}
  \centering
  \includegraphics[width=8cm, height=5cm]{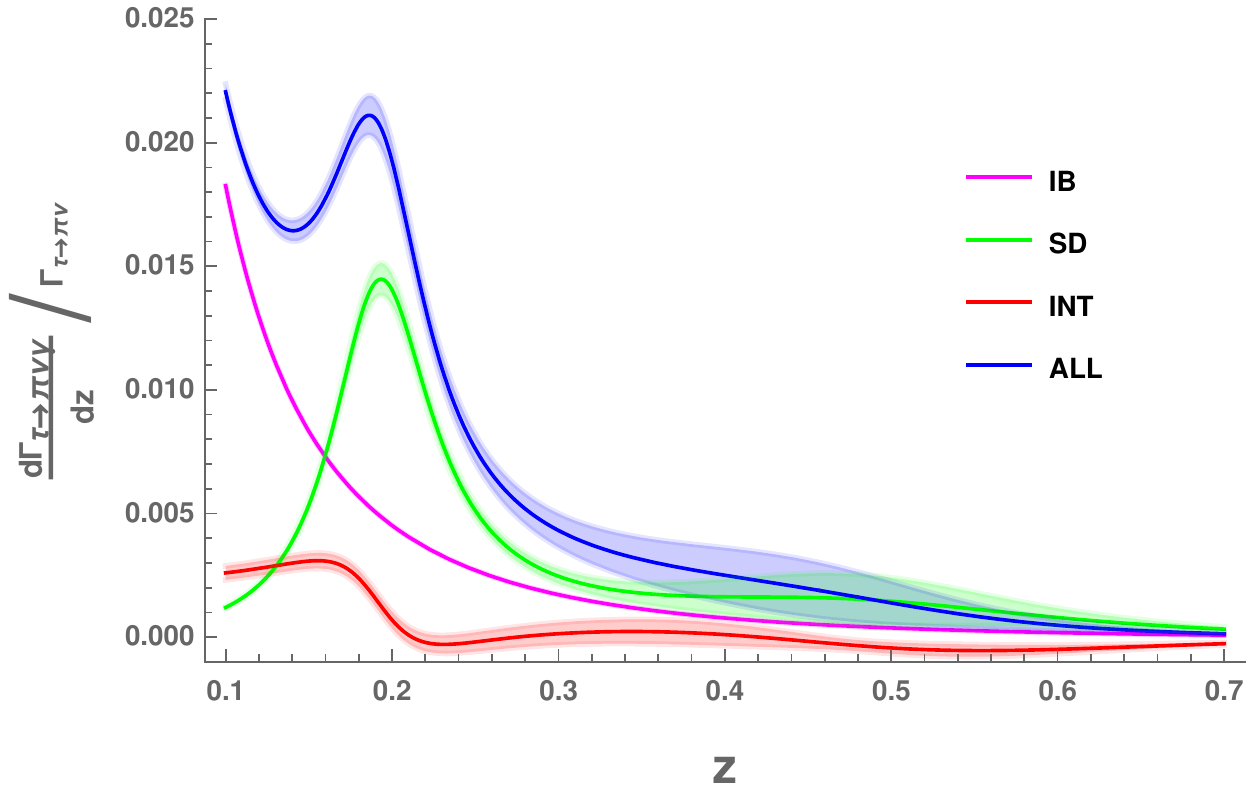}
  \caption{}
\end{subfigure}%
\begin{subfigure}{.5\textwidth}
  \centering
  \includegraphics[width=8cm, height=5cm]{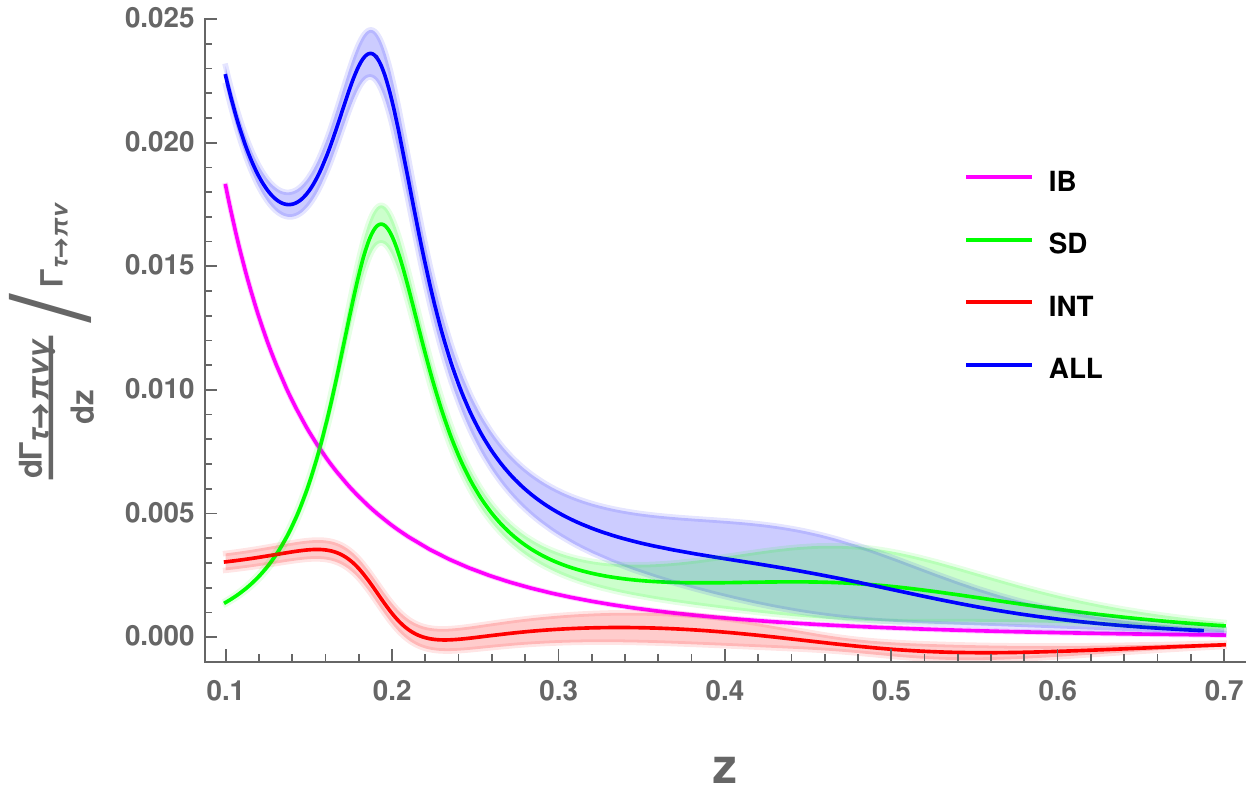}
  \caption{}
\end{subfigure}
\caption{The invariant mass spectrum of $\pi-\gamma$ system for radiative tau decay is shown here considering (a) asymptotic and (b) CZ pion distribution amplitude. The contributions from the IB (magenta), SD (green) and the interference (red) of the two is also shown. The shaded region shows the uncertainties.}
\label{all}
\end{figure}

The analytic expressions for the vector and axial-vector form factors calculated using LCSR are given in Eq.(\ref{FVF}) and Eq.(\ref{FAF}). Both these form factors have the asymptotic $\frac{1}{t}$ dependence on the invariant mass squared, $t$ of the photon pion system, as expected from QCD in the perturbative (asymptotic) regime. We have used two forms of pion distribution amplitude; the asymptotic form and the CZ form as given in Eq.(\ref{phia}) and Eq.(\ref{phiz}), respectively. The structure dependent parameter defined in Eq.(\ref{sdp}) is also calculated using both the forms for pion distribution amplitudes. The values of the various parameters used for the numerical computation are collected in Appendix-\ref{appendixB}. The form factors depend on the value of the Borel parameter, $M$, and hence also the structure dependent parameter, $\gamma$. Fig.(\ref{sdpp}) shows the variation of $F_A^{(\pi)}(0)$, $F_V^{(\pi)}(0)$ and SDP ($\gamma$)  with the variation in the value of $M$. The variation of the observables with $M$ dictates the model dependence here. As can be seen from the plot, all the observables are quite stable in the chosen Borel window. The value of $\gamma$ for $M=3.35$ GeV is 0.469 (using CZ distribution amplitude) which matches well, including the sign, with the experimental value of $\gamma$ obtained from the radiative pion decay \cite{pdg}. \\
Further, we calculate the decay width contribution for the radiative tau decay using $M=3.35$ GeV and the form factors given in Eq.(\ref{FVF}) and Eq.(\ref{FAF}). 
The differential decay rate for the radiative tau decay is given by,
\begin{equation}
 d\Gamma(\tau^-\rightarrow \pi^- \nu_\tau \gamma) = \frac{1}{512 \pi^5}E_\tau \delta^{(4)}(k+p_2+p_3-p_1)\overline{|\mathcal{A}|^2} \frac{d^3 k d^3p_2 d^3 p_3}{E_\gamma E_\pi E_\nu}
 \label{dw}
\end{equation}
where, $E_\tau, E_\pi, E_\gamma, E_\nu$ are the energies of tau-lepton, pion, photon and neutrino, respectively. $\overline{|\mathcal{A}|^2}$ is the spin averaged square of the amplitude which has been calculated in Section-\ref{Amplitude}.\\
In terms of the functions used in Eq.(\ref{final}), %$\overline{|\mathcal{A}|^2}$ will be ,
\begin{equation}
    \overline{|\mathcal{A}|^2} = \overline{|\mathcal{A}_{IB}|^2} + \overline{|\mathcal{A}_{SD}|^2}+ 2\mathcal{R}e\{\overline{\mathcal{A}_{IB}^*\mathcal{A}_{SD}}\}
\end{equation}
where, $\overline{|\mathcal{A}_{SD}|^2}= \overline{|\mathcal{A}_{A}|^2}+\overline{|\mathcal{A}_{V}|^2}+ 2\mathcal{R}e\{\overline{\mathcal{A}_{A}^*\mathcal{A}_{V}}\}$.\\
The kinematical details to compute the decay rate can be found in Appendix-\ref{kinematics}.\\
\begin{figure}[h]
     \centering
     \includegraphics[width=9.5cm, height=6.5cm]{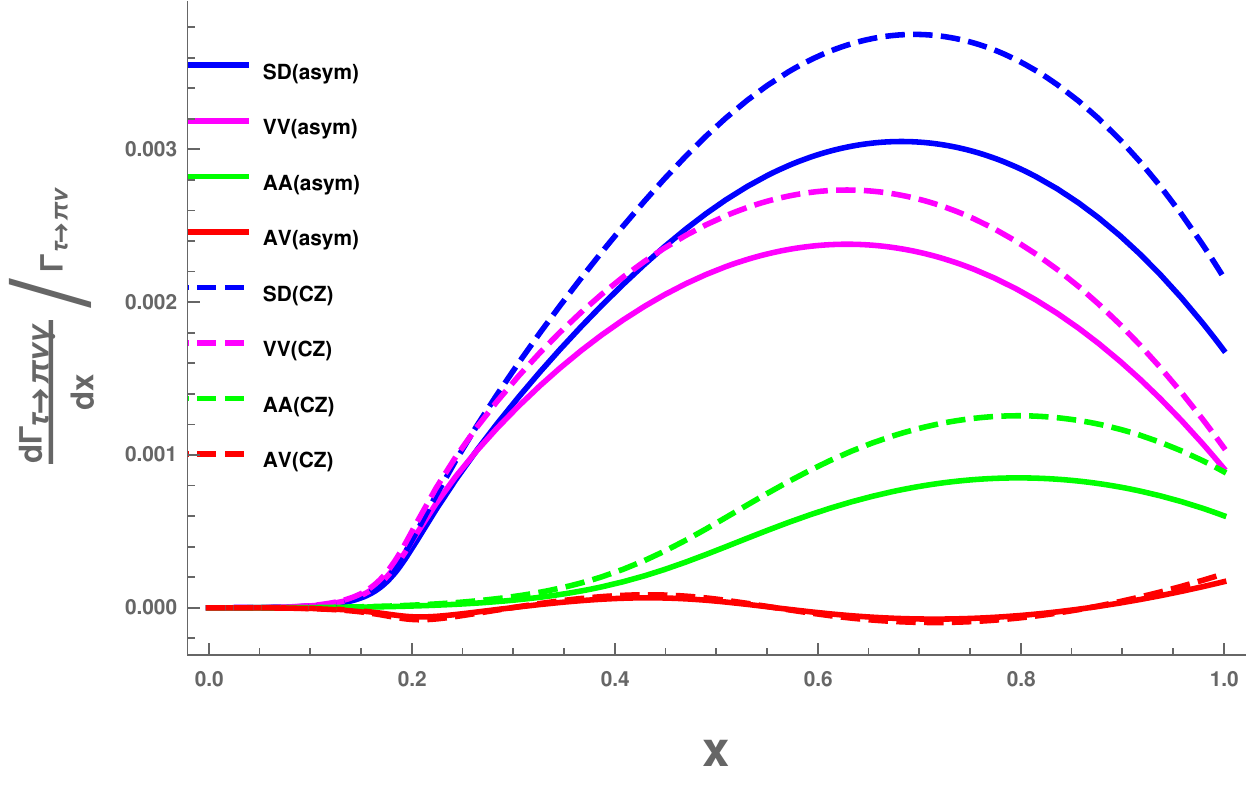}
     \caption{The total Structure Dependent Contribution (blue) to the photon spectrum is shown along with the individual contributions from the vector (magenta), axial vector (green) and the interference (red) of the two are also shown for the two distribution amplitudes.  Solid lines are for asymptotic distribution amplitude while dashed ones are for Chernyak-Zitnisky distribution amplitude.}
     \label{SD_pion}
 \end{figure} 
The  structure dependent contribution to the photon spectrum is shown in Fig.(\ref{SD_pion}) using both forms of pion distribution amplitudes. 
\begin{figure}[h]
     \centering
     \includegraphics[width=9.5cm, height=6.5cm]{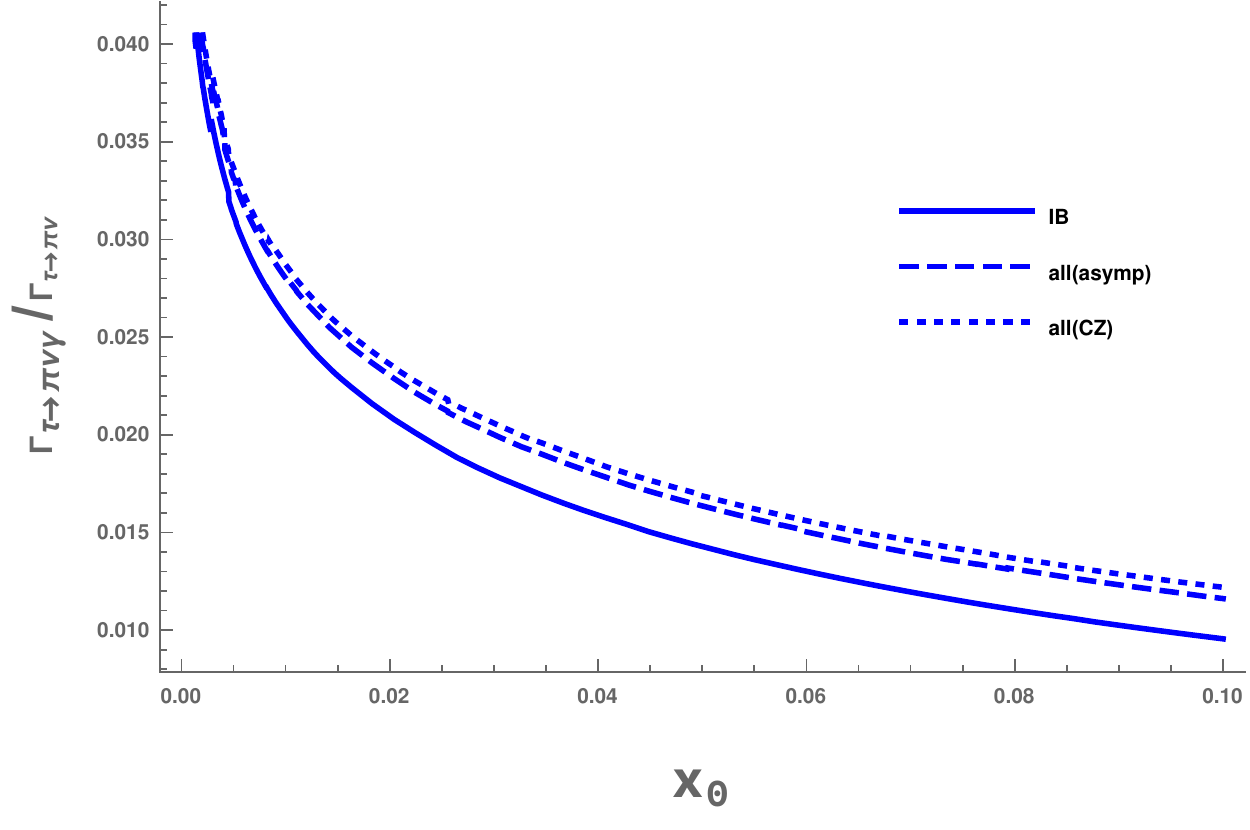}
     \caption{The dependence of the IB (solid) contribution on the minimum energy threshold of the photon is shown here. Along with that, the same dependence for total decay width including form factors using asymptotic (dashed) and CZ (dotted) pion distribution amplitude is also shown.}
     \label{thdep}
 \end{figure}
The IB contribution suffers from the infrared divergences which can be taken care of by putting a threshold on the photon energy. Fig.(\ref{thdep}) shows the threshold energy dependence of the IB contribution as well as the full decay width of the radiative tau decay. The SD contribution is free from any kind of divergences.\\
$F_A^{(\pi)}(t)$ gets contribution from $a_1$ meson while $F_V^{(\pi)}(t)$ from $\rho$ (and $\omega$) meson.
 \begin{figure}
\centering
\begin{subfigure}{.5\textwidth}
  \centering
  \includegraphics[width=8cm, height=5cm]{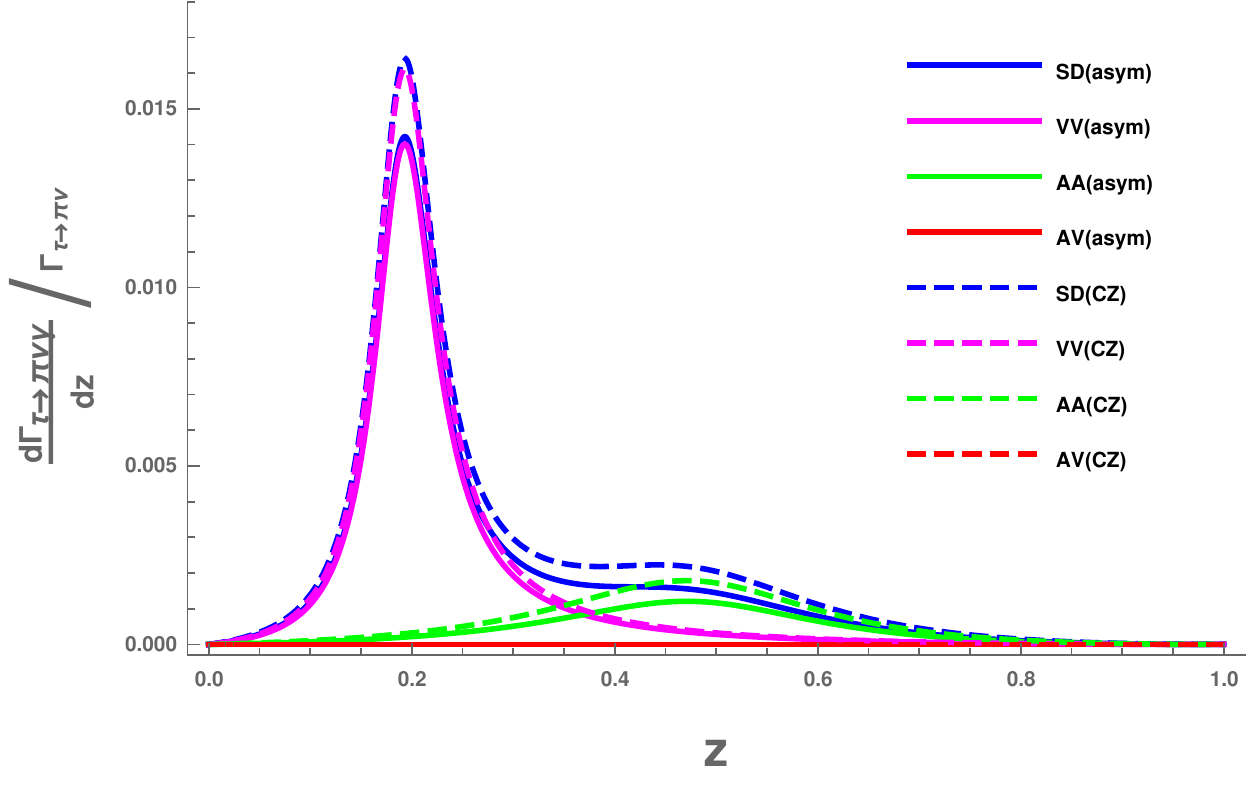}
  \caption{}
\end{subfigure}%
\begin{subfigure}{.5\textwidth}
  \centering
  \includegraphics[width=8cm, height=5cm]{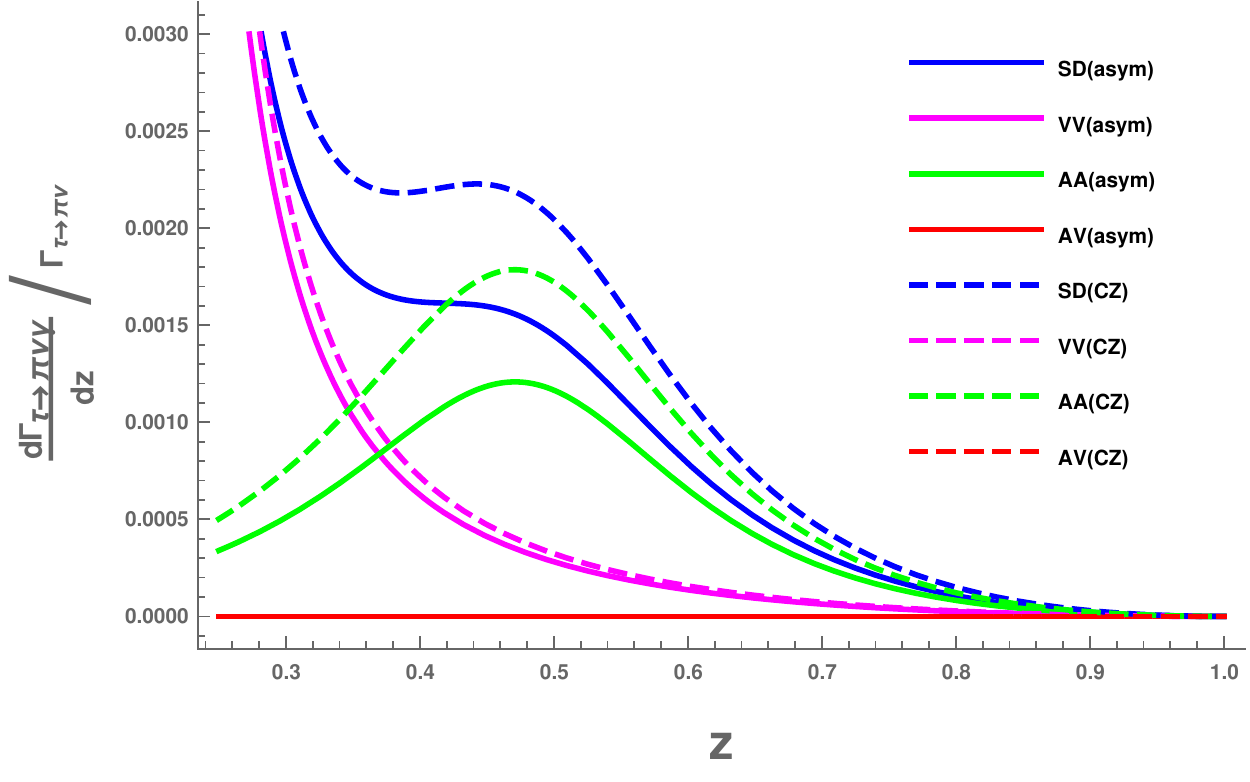}
  \caption{}
\end{subfigure}
\caption{(a): The Structure Dependent contribution (blue) to the invariant mass spectrum of $\pi-\gamma$ system is shown here for asymptotic (solid) and Chernyak-Zhitnisky (dashed) pion distribution amplitudes. The contribution from the vector (magenta), axial vector (green) and the interference (red) of the two is also shown. (b): Zoomed in version of (a).}
\label{SD_pigamma}
\end{figure}
Fig.(\ref{SD_pigamma}) shows the SD contribution to the invariant mass spectrum of $\pi-\gamma$ system. The higher and sharper peak corresponds on the contribution coming from the vector mesons while the shorter and broader peak corresponds to the axial vector contribution. 
The vector contribution to the total decay width dominates over the axial-vector contribution.\\
As $\rho$ and $a_1$-mesons are not very narrow, the effect of $t$ dependence of the widths is also studied using the 
prescription provided in \cite{Kuhn:1990ad}.
The $t$ dependence of $\Gamma_\rho$ does not have significant effect as it is not that wide while 
the effect of $\Gamma_{a_1}$ is clearly visible as one can see from Fig.(\ref{q2}). 
The explicit forms of $t$ dependence of the decay widths are collected in Appendix-\ref{appendixA}. We have also computed the 
effect of decay width of $a_1$-meson $\Gamma_{a_1}$, as it has huge uncertainty, and found that the decay width of radiative
tau decay decreases with an increase in $\Gamma_{a_1}$. The results reported here are calculated using $\Gamma_{a_1}= 425$ MeV.\\
 Fig.(\ref{all}) represents all the contributions to the invariant mass spectrum of the $\pi-\gamma$ system. The IB contribution dominates at the low photon energy for which we have used the minimum energy threshold of $50$ MeV.\\
  \begin{figure}
\centering
\begin{subfigure}{.5\textwidth}
  \centering
  \includegraphics[width=8cm, height=5cm]{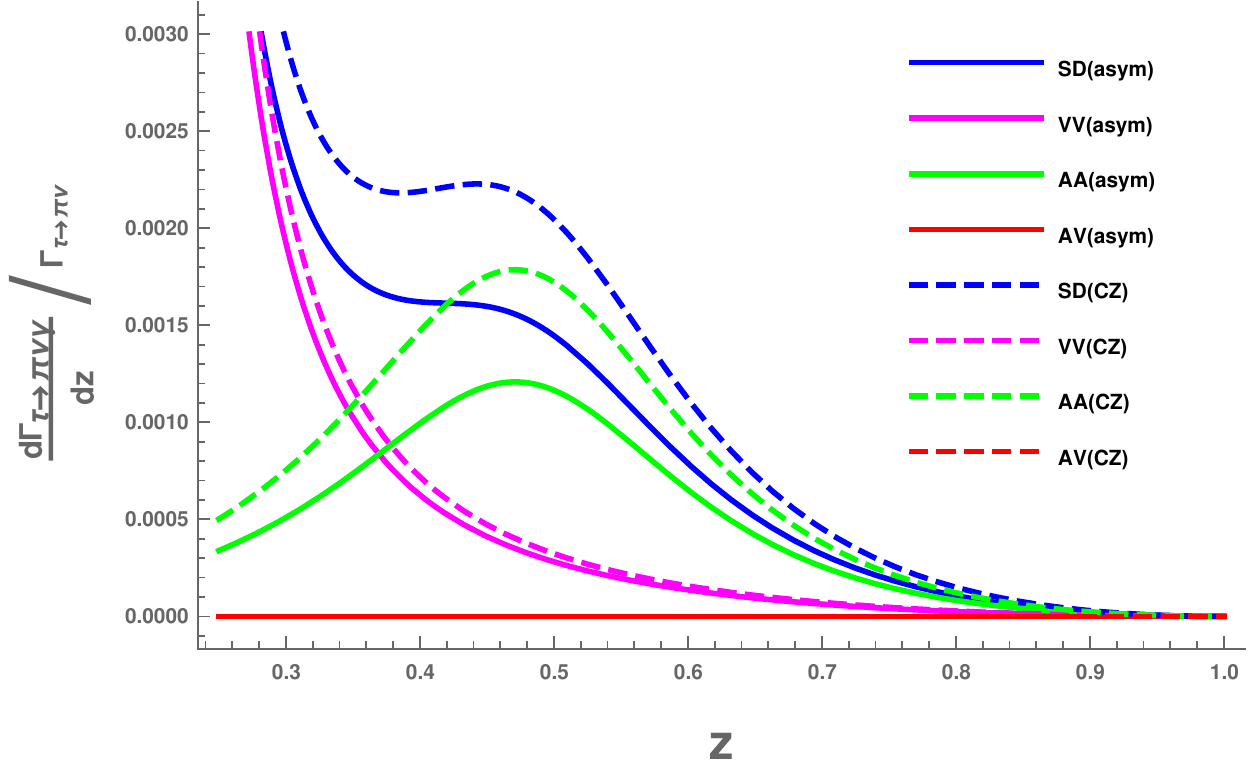}
  \caption{}
\end{subfigure}%
\begin{subfigure}{.5\textwidth}
  \centering
  \includegraphics[width=8cm, height=5cm]{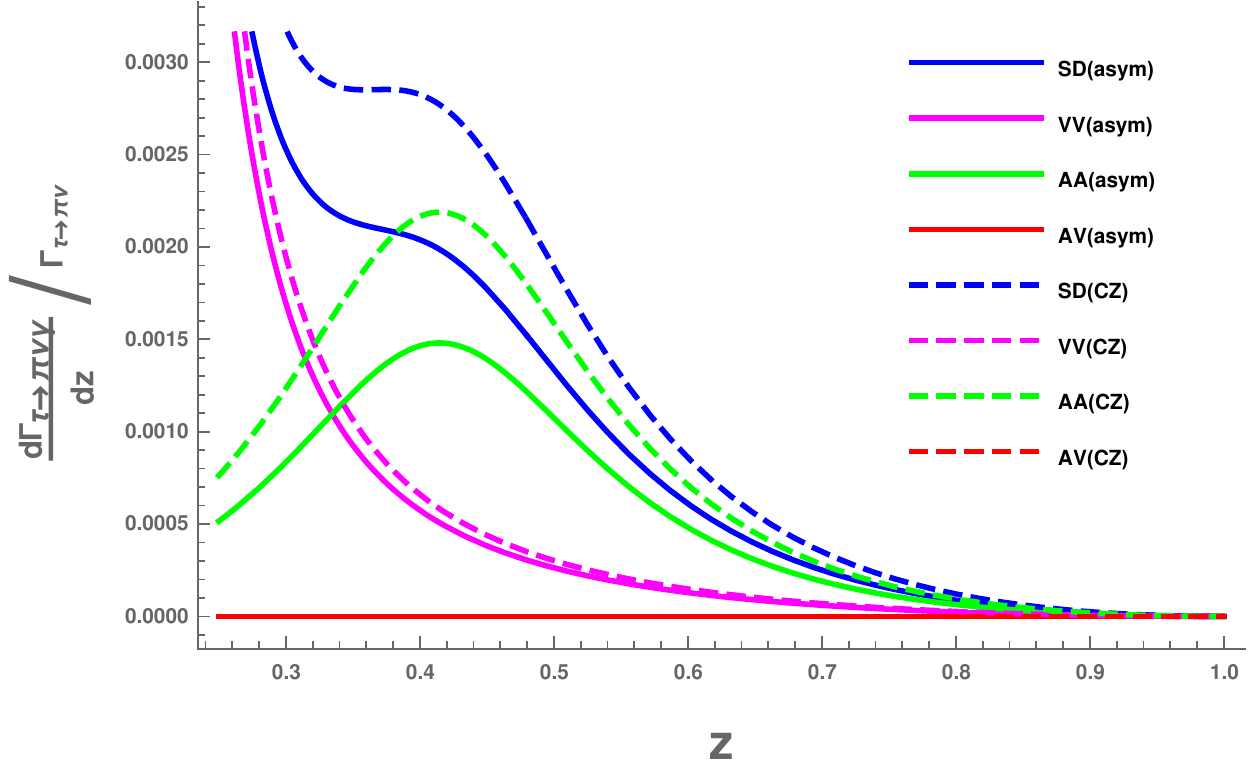}
  \caption{}
\end{subfigure}
\caption{The SD contribution (blue) considering (a) $\Gamma_\rho$ and $\Gamma_{a_1}$ to be constant and (b) the $t$ dependence of $\Gamma_\rho$ and $\Gamma_{a_1}$ is shown here for asymptotic (solid) and Chernyak-Zhitnisky (dashed) pion distribution amplitudes. The contribution from the vector (magenta), axial vector (green) and the interference (red) of the two is also shown.}
\label{q2}
\end{figure}
  %\begin{figure}
   %  \centering
    % \includegraphics[width=9.5cm, height=6.5cm]{All_pion_spectrum3.pdf}
     %\caption{The invariant mass spectrum of $\pi-\gamma$ system for radiative tau decay is shown here. The contributions from the IB (magenta), SD (green) and the interference (red) of the two is also shown.}
   %  \label{all}
% \end{figure}
 After integrating over the full phase space and applying a energy threshold of 50 MeV for the IB contribution, we get the following values for the different contributions to the decay width (normalised to the non-radiative decay width Eq.(\ref{nonradiativewidth}) ie $\bar{\Gamma} = \Gamma(\tau\to\pi\nu\gamma)/\Gamma(\tau\to\pi\nu)$):\\ 
 (i) Asymptotic pion distribution amplitude:
 $$\bar{\Gamma}_{IB}= 1.36\times 10^{-2}, \hspace{0.5 cm} \bar{\Gamma}_{VV}=  (1.47\pm0.06)\times 10^{-3}, \hspace{0.5 cm} \bar{\Gamma}_{AA}=  (3.97\pm2.45)\times 10^{-4}, \hspace{0.5cm} \bar{\Gamma}_{AV} \approx 0$$
 $$ \bar{\Gamma}_{SD}=  (1.87\pm0.30)\times 10^{-3}, \hspace{0.5 cm} \bar{\Gamma}_{int}= (3.82\pm2.14)\times 10^{-4}, \hspace{0.5 cm} \bar{\Gamma}_{all}= (1.56\pm0.04)\times 10^{-2}$$
(ii) CZ pion distribution amplitude: 
 $$\bar{\Gamma}_{IB}= 1.36\times 10^{-2}, \hspace{0.5 cm} \bar{\Gamma}_{VV}=  (1.70\pm0.07)\times 10^{-3}, \hspace{0.5 cm} \bar{\Gamma}_{AA}=  (5.91\pm3.62)\times 10^{-4}, \hspace{0.5cm} \bar{\Gamma}_{AV} \approx 0$$
 $$ \bar{\Gamma}_{SD}= ( 2.29\pm0.43)\times 10^{-3}, \hspace{0.5 cm} \bar{\Gamma}_{int}= (4.90\pm2.60)\times 10^{-4}, \hspace{0.5 cm} \bar{\Gamma}_{all}= (1.61\pm0.06)\times 10^{-2}$$
 Since we consider radiative rate normalised to the non-radiative one, the uncertainty in IB contribution is negligible compared to the SD contribution which dominates the error budget therefore no uncertainty is shown for the IB part. The final uncertainties are about $10\%$. From the above it is evident that there is a dependence on the form of the distribution amplitude chosen to evaluate these form factors. However, the difference is not too large, which is reassuring.\\
 Having obtained detailed predictions for the pion in the final state, it is also instructive to have an estimate of the decay width for the kaon in the final state. Again, normalising to the appropriate non-radiative width, and employing the asymptotic distribution amplitude (keeping the Borel parameter, $M=3.35$ GeV), we get
 \begin{equation}  
 \bar{\Gamma}^K = \Gamma(\tau\to K\nu\gamma)/\Gamma(\tau\to K\nu) \sim 8\times 10^{-3}
 \end{equation}

 This (approriately normalised) rate is roughly half of that for the pion.  

\section{Discussion and Conclusions}
\label{discussion}
In the present paper, we have provided detailed predictions for the rate and photon spectrum for the process 
$\tau^- \rightarrow \pi^- \nu_\tau \gamma$. Employing Ward identity from the beginning, the amplitude was written so as 
to include the contact term which is necessitated by gauge invariance. The decay involves two time like form factors. 
These have been calculated in the present work employing the Light Cone Sum Rules, to twist-2 accuracy. The form factors, 
which automatically via the dispersion relations, encode the contributions from the vector and axial-vector mesons,
have the right asymptotic behaviour expected from perturbative QCD. The ratio of the axial-vector to vector form factor at 
zero momentum transfer defines the pion structure dependent parameter, $\gamma$. Our evaluation of this parameter, along with the 
sign, matches very well with the experimental value obtained from $\pi\to\ell\nu\gamma$, where the relevant pion-photon form factors, 
unlike the present case, are space like. The obtained values for the normalised rate and the photon spectrum are similar to those 
obtained in \cite{roig}. This provides a cross-check on the theoretical predictions employing a totally different method for 
computing the non-perturbative quantities. We have also provided an estimate for the appropriately normalised rate with kaon
in the final state instead of a pion. This normalised rate is approximately half of that for the pion. The present
study  employed distribution amplitudes to twist-2 accuracy. 
The uncertainties reported here are the uncertainties associated with the uncertainities of the various parameters used. 
There will be further uncertainties associated with quark hadron duality approximation, and higher twist and hight order contributions.
The pion is considered to be massless here. The effect of such an assumption is less than 1$\%$ on the values of the form factors.
 The uncertainties associated with quark hadron duality violation, like in inclusive tau decays are expected to be at $10\%$ level,
 and can be calculated in 
 a particular model to parametrise the spectral density. Precise calculations of these duality violations is indeed an important
 missing piece but is out of the scope of present work. It would be interesting to consider both higher 
 twist contributions as well as contributions higher order in $\alpha_s$. These can have a significant impact on the phenomenology of 
 radiative one meson tau decays.

\begin{appendix}
\numberwithin{equation}{section}
\section{Conventions, Definitions and Identities}
\label{appendixA}
Here, we are reporting the various conventions and definitions used for the sake of completeness,
\begin{enumerate}
    \item The matrix element of the pion is defined as;
    \begin{equation}
       \left<\pi^-(p_2)|(\bar d \gamma^\mu (1-\gamma_5) u )|0\right> = i f_\pi p_2^\mu 
    \end{equation}
where, $f_\pi$ is the pion decay constant.
\item The outgoing photon state can be obtained by the use of creation operator on the vacuum which results into,
\begin{equation}
    \left<\nu_\tau \gamma| \bar \nu_\tau \gamma_\mu (1-\gamma_5)|\tau^-\right> = -ie \epsilon_\mu^* \int d^4 x e^{i k x } \left<\nu_\tau |T\{j_{em}^\alpha (x) \bar \nu_\tau \Gamma_\mu \tau(0)\}|\tau^-\right>
\end{equation}
where, $j_{em}^\alpha(x) = Q_\psi \bar\psi(x)\gamma^\alpha \psi(x) = - \bar{\tau}\gamma^\alpha \tau + Q_u \bar{u}\gamma^\alpha u + Q_d \bar{d}\gamma^\alpha d$ is the electromagnetic current. $Q_u$ and $Q_d$ are the electromagnetic charges of $u$ and $d$ quarks, respectively in the units of $e$.
\item The commutator of the electromagnetic charge operator and electroweak current of the pion is given by,
\begin{equation}
    \left[j_{em}^0(x), \bar d \Gamma^\mu u(0)\right] = -Q_u \delta^3 (x) \bar d(0)\Gamma^\mu u(x) +Q_d \partial^3 (x) \bar d(x)\Gamma^\mu u(0)
\end{equation}
\item The propagator of the massless fermions in position space is given by,
\begin{equation}
    iS_0 (x)= \left<0|T\left\{u(x)\bar u(0)\right\}|0\right>= \frac{i\slashed{x}}{2\pi^2x^4} = -\left<0|T\left\{u(0)\bar u(x)\right\}|0\right>
\end{equation}
\item $\gamma_\mu \gamma_\beta \gamma_\alpha = g_{\mu\beta}\gamma_\beta - g_{\mu\alpha}\gamma_\beta + g_{\beta\alpha}\gamma_\mu - i\epsilon_{\mu\beta\alpha\rho}\gamma^\rho \gamma_5$
\item The leading order expansion (twist-2) of the matrix element $\left<\pi^-(p_2)|\bar d(y) \gamma_\mu \gamma_5 u(x)|0\right>$ in the light cone limit ($x^2=0$) is given by,
\begin{equation}
    \left<\pi^-(p_2)|\bar d(y) \gamma_\mu \gamma_5 u(x)|0\right> = if_\pi p_{2\mu} \int_0^1 du e^{i(up_2y+ \bar u p_2 x)} \phi(u,\mu)
\end{equation}
where, $\bar u = 1-u$ and $\phi(u,\mu)$ is pion distribution amplitude of twist-2. 
\item The matrix elements of $\rho$ and $a_1$ mesons are defined as,
\begin{align}
    &\left<V(p_2+k)|\bar d \gamma_\mu u|0\right> = -i m_V f_V \epsilon_\mu^{(V)*}\\
    &\left<\pi^-(p_2)|j_{em}^\alpha(x)|\rho(p_2+k)\right> = \epsilon^{\alpha \lambda\beta\nu} \epsilon_\lambda^{(\rho)}p_{2\beta}k_\nu F_{\rho\pi}(k^2)\\
    &\left<\pi^-(p_2)|j_{em}^\mu(x)|a_1(p_2+k)\right> = \left[(2p_2-k).k g^{\mu\lambda} - (2p_2-k)^\mu k^\lambda\right]\epsilon_\lambda^{(a_1)*} G_{a_1 \pi}(k^2)
\end{align}
 where, $V$ can be $\rho$ or $a_1$-meson, $m_V$ and $f_V$ are the mass and decay constant of $V$-meson, respectively. $\epsilon_\lambda^{(\rho)}$ and $\epsilon_\lambda^{(a_1)*}$ are the polarisation vectors of $\rho$ and $a_1$-meson, respectively. $ F_{\rho\pi}(k^2)$ and $G_{a_1 \pi}(k^2)$ are the scalar functions of $k^2$ which contains the information of $\rho\rightarrow\pi$ and $a_1\rightarrow\pi$ transitions, respectively.
 \item The sum over polarisation of $\rho$ or $a_1$-meson is given by,
\begin{equation}
\epsilon_\lambda^{(V)}\epsilon_\nu^{(V)*} = -g_{\lambda\nu}+\frac{(p_2+k)_\lambda(p_2+k)_\nu}{m_{V}^2}
\end{equation}
\item { The $t$-dependence of the deacy widths of $\rho$ and $a_1$ mesons are given by \cite{Kuhn:1990ad},
\begin{equation}
    \Gamma_\rho(t) = \Gamma_\rho \frac{m_\rho^2}{p_\rho^3}\frac{p^3}{t}
\end{equation}
with, $2p = (t-4m_\pi^2)^{1/2}$ and $2p_\rho = (m_\rho^2-4 m_\pi^2)^{1/2}$.
\begin{equation}
    \Gamma_{a_1}(t) = \frac{m_{a_1} \Gamma_{a_1}}{\sqrt{t}} \frac{g(t)}{g(m_{a_1}^2)} 
\end{equation}
with,
\[
 g(t) =
\begin{cases}
  4.1(t-9m_\pi^2)^3(1-3.3(t-9m_\pi^2)+5.8(t-9m_\pi^2)^2) & \text{if  \hspace{0.3cm}  $t< (m_\rho+m_\pi)^2$} \\
  t(1.623 + \frac{10.38}{t} - \frac{9.38}{t^2}+\frac{0.65}{t^3}) & \text{else}
\end{cases}
\]

}

\end{enumerate}

\section{Values of parameters used}
\label{appendixB}
{
Here, we tabulate the values of the various parameters used for numerical calculation. 
\begin{center}
\begin{tabular}{ |c||c|c|c|c| } 
 \hline
\textbf{S.No.} &\textbf{Parameter} \hspace{2cm} & \textbf{Symbol}\hspace{1cm} & \textbf{ Value} \hspace{2cm}  \\ 
 \hline\hline
 \textbf{1.}&Fine structure constant & $\alpha$& ${\frac{1}{133.6}}$  \\ 
 \hline
 \textbf{2.}&Fermi's Constant & $G_F$ & $1.166 \times 10^{-5} \text{ GeV}^{-2}$\cite{pdg} \\ 
 \hline
\textbf{3.}&Mass of $\tau$-lepton & $m_\tau$& $(1776.86\pm0.12)$ MeV \cite{pdg}  \\ 
 \hline
 \textbf{4.}&Pion decay constant & $f_\pi$& $(130.41\pm0.23)$ MeV   \\ 
 \hline
 \textbf{5.}&CKM Matrix element & $V_{ud}$& { $(0.9745\pm0.0001)$} \cite{pdg} \\ 
 \hline
 \textbf{6.}&Mass of $\rho$-meson & $m_\rho$& $(775.26\pm0.25)$ MeV \cite{pdg} \\ 
 \hline
 \textbf{7.}&Decay width of $\rho$-meson & $\Gamma_\rho$& $(149.1\pm0.8)$ MeV  \cite{pdg} \\ 
 \hline
 \textbf{8.}&Mass of $a_1$-meson & $m_{a_1}$& $(1230\pm40)$ MeV  \cite{pdg} \\
 \hline
 \textbf{9.}&Decay width of $a_1$-meson & $\Gamma_{a_1}$& $(425\pm175)$ MeV  \cite{pdg} \\ 
 \hline
 \textbf{10.}&Vector form factor & $F_V^{(\pi)}(0)$& $0.0254\pm 0.0017$  \cite{pdg} \\ 
 \hline
 \textbf{11.}&Axial-vector form factor & $F_A^{(\pi)}(0)$& $0.0119\pm 0.0001$ \cite{pdg}  \\ 
 \hline
 \textbf{12.}&$\alpha_s(1 GeV)$& $\alpha_s(1 GeV)$& $\sim 0.7$   \\ 
 \hline
 \textbf{13.}&$\alpha_s(m_\tau)$& $\alpha_s(m_\tau)$& $0.325$   \\ 
 \hline
 \textbf{14.}&$a_2(1 GeV)$& $a_2(1 GeV)$& $0.12$  \\ 
 \hline
\end{tabular}
\end{center}
{The value of the fine structure constant is taken at the scale $m_\tau$ and the decay width of $a_1$ meson is taken to the central value of the range given in \cite{pdg}.}
}\section{Kinematics and Decay width}
\label{kinematics}
The differential decay width can be written as a sum of different components \cite{decker}:  $\Gamma_{IB}$ coming from  $\overline{|\mathcal{A}_{IB}|}^2$, $\Gamma_{SD}$ coming from  $\overline{|\mathcal{A}_{SD}|}^2$ and $\Gamma_{int}$ coming from $2 \overline {\mathcal{R}e(\mathcal{A}_{IB}^* \mathcal{A}_{SD})}$. $\Gamma_{SD}$ is further divided into three parts: $ \Gamma_{VV}$  coming from  $\overline{|\mathcal{A}_{v}|}^2$, $ \Gamma_{AA}$  coming from  $\overline{|\mathcal{A}_{A}|}^2$ and $ \Gamma_{AV}$  coming from  $2 \overline {\mathcal{R}e(\mathcal{A}_{V} \mathcal{A}_{A}^*)}$. 
\begin{align}
 \nonumber & \Gamma_{all} = \Gamma_{IB}+\Gamma_{int}+\Gamma_{SD}, \\ \nonumber &
 \Gamma_{SD} =\Gamma_{VV}+ \Gamma_{AV} +\Gamma_{AA} ,
 \\ &
 \Gamma_{int} =\Gamma_{IB-A}+ \Gamma_{IB-V}  .
 \label{divide}
 \end{align}
For convenience, we use the dimensionless variables $x$ and $y$ defined as,
%(as used in \cite{roig} and references therein) define as,
\begin{equation}
 x = \frac{2p_1.k}{m_\tau^2}, \hspace{0.5cm} y = \frac{2p_1.p_2}{m_\tau^2}.
\end{equation}
In the rest frame of tau, $x$ and $y$ are simply the energies of photon and pion respectively in units of $\frac{m_\tau}{2}$. The kinematical boundaries of $x$ and $y$ are given by,
\begin{equation}
 0 \leq x \leq 1-r_p^2, \hspace{0.5cm} 1-x+\frac{r_p^2}{1-x} \leq y \leq 1+r_p^2
\end{equation}
where, $r_p^2 = \frac{m_\pi^2}{m_\tau^2}$. We have considered pion to be massless for form factor calculations and hence we will use $r_p \rightarrow 0$ in our final answers.\\
The variable $t$, the invariant mass square of the pion-photon system, can be written in terms of $x$ and $y$ as
\begin{equation}
 t = P^2 = (p_2+k)^2 = m_\tau^2(x+y-1) \hspace{0.5cm} \implies P.k = \frac{m_\tau^2}{2}(x+y-1-r_p^2).
\end{equation}

In terms of variables $x$ and $y$, the differential decay width in the rest frame of tau is,
\begin{equation}
 \frac{d^2\Gamma}{dxdy} = \frac{m_\tau}{256 \pi^3}\overline{|\mathcal{A}|^2},
\end{equation}
where different contribution to the differential decay width are (calculated using FeynCalc \cite{Shtabovenko:2020gxv}):
\begin{align}
    & \nonumber \frac{d^2\Gamma_{IB}}{dxdy} = \frac{\alpha}{2\pi} f_{IB}(x,y,r_p^2) \frac{\Gamma_{\tau^- \rightarrow \pi^-\nu_\tau}}{(1-r_p^2)^2},\\
    & \nonumber \frac{d^2\Gamma_{SD}}{dxdy} = \frac{\alpha}{8\pi}\frac{m_\tau^4}{f_\pi^2}\left\{|F_V^{(\pi)}|^2f_{VV}(x,y,r_p^2) + 2\mathcal{R}e(F_A^{(\pi)*}F_V^{(\pi)}) f_{AV}(x,y,r_p^2)  + |F_A^{(\pi)}|^2f_{AA}(x,y,r_p^2) \right\} \frac{\Gamma_{\tau^- \rightarrow \pi^-\nu_\tau}}{(1-r_p^2)^2},\\
&    \frac{d^2\Gamma_{int}}{dxdy} = \frac{\alpha}{2\pi}\frac{m_\tau^2}{f_\pi} \left[f_{IB-V}(x,y,r_p^2) \mathcal{R}e(F_V^{(\pi)}) + f_{IB-A}(x,y,r_p^2) \mathcal{R}e(F_A^{(\pi)})\right] \frac{\Gamma_{\tau^- \rightarrow \pi^-\nu_\tau}}{(1-r_p^2)^2},
\end{align}
 with $\alpha = \frac{e^2}{4\pi}$, being the fine structure constant,
 \begin{align}
     & \nonumber f_{IB}(x,y,r_p^2)= \frac{\left[r_p^4(x+2)-2r_p^2(x+y)+(x+y-1)(2-3x+x^2+xy)\right](r_p^2-y+1)}{(r_p^2-x-y+1)^2x^2}, \\
     & \nonumber  f_{VV}(x,y,r_p^2) = -\left[r_p^4(x+y)+2r_p^2(1-y)(x+y)+(x+y-1)(-x+x^2-y+y^2)\right],\\
     & \nonumber  f_{AV}(x,y,r_p^2) = -\left[r_p^2(x+y)+(1-x-y)(y-x)\right](r_p^2-x-y+1),\\
     & \nonumber f_{AA}(x,y,r_p^2) = f_{VV}(x,y,r_p^2),\\
     & \nonumber f_{IB-V}(x,y,r_p^2) = -\frac{(r_p^2-x-y+1)(r_p^2-y+1)}{x}, \\
     & f_{IB-A}(x,y,r_p^2) = -\frac{\left[r_p^4-2r_p^2(x+y)+ (1-x+y)(x+y-1)\right](r_p^2-y+1)}{(r_p^2-x-y+1)x},
 \end{align}
 and $\Gamma_{\tau^- \rightarrow \pi^-\nu_\tau}$ is the non-radiative decay width given by,
 \begin{equation}
     \Gamma_{\tau^- \rightarrow \pi^-\nu_\tau} = \frac{G_F^2|V_{ud}|^2f_\pi^2}{8\pi} m_\tau^3(1-r_p^2)^2.
     \label{nonradiativewidth}
 \end{equation}
The photon spectrum is obtained by integrating over $y$. Integration over $x$ will give the total decay width for radiative tau decay. The IB contribution has the infrared divergences which can be fixed by putting a threshold on the minimum energy of the emitted photon. SD contribution does not face any such divergence and hence can be integrated over the full phase space.
\begin{equation}
    \Gamma(\tau^-\rightarrow \pi^- \nu_\tau\gamma)= \int_{x_0}^{1-r_p^2}dx \int_{1-x+\frac{r_p^2}{1-x}}^{1+r_p^2} dy \frac{d^2\Gamma}{dxdy}
\end{equation}
where, $x_0$ is the minimum energy cut for the photon energy in unit of $\frac{m_\tau}{2}$.\\
To get the invariant mass spectrum of $\pi\gamma$ system, define another dimensionless variable $z$ (as used in ref-\cite{decker}) as,
\begin{equation}
 z = \frac{t}{m_\tau^2} = x+y-1.
\end{equation}
The kinematical boundaries for the new variable are:
\begin{equation}
 z-r_p^2 \leq x \leq 1-\frac{r_p^2}{z}, \hspace{0.5cm} r_p^2\leq z\leq 1.
\end{equation}
The $\pi\gamma$ spectrum can be obtained by substituting $y$ in terms of $z$ in $\frac{d^2\Gamma}{dxdy}$ and integrating it over $x$, i.e,
\begin{equation}
 \frac{d\Gamma}{dz}= \int_{z-r_p^2}^{1-\frac{r_p^2}{z}} dx \frac{d^2\Gamma}{dxdy}(x,y=z-x+1).
\end{equation}
\end{appendix}

\bibliography{Radiative_tau}{}
\bibliographystyle{unsrt}
%\nocite{*}
\end{document}